# Emergent Non-Hermitian Topology in Multi-Robot Network

Jielong Zhang[1,5], Guiju Duan[1,2,3,5], Tinggui Chen[1,5], Shengjie Zheng[1], Bozheng Xue[4], Baizhan Xia[1,4*]

1 State Key Laboratory of Advanced Design and Manufacturing for Vehicle Body, Hunan University, Changsha, Hunan, People's Republic of China, 410082;

2 AMOLF, 1098 XG Amsterdam, The Netherlands

3 Huygens-Kamerlingh Onnes Laboratory, Leiden, The Netherlands

4 College of Mechanical and Vehicle Engineering, Hunan University, Changsha, Hunan, People's Republic of China, 410082

5 These authors contributed equally: Jielong Zhang, Guiju Duan, Tinggui Chen.

*Email: xiabz2013@hnu.edu.cn

**Non-Hermitian (NH) topology has been extensively explored in wave and matter systems, typically relying on the routing of complex, non-reciprocal couplings in physical space. This work demonstrates the experimental realization of programmable NH topological phases within decentralized multi-robot networks. By digitally programming non-reciprocal interaction rules and establishing real-time state exchange among active robots, we observe emergent topological zero modes (TZMs) and NH skin effects in synthetic lattices spanning one to three dimensions. Dynamically tailoring non-reciprocal parameters enables the precise morphing of TZMs between localized and delocalized states, establishing a versatile framework for topological mode engineering across dimensionalities. This platform establishes multi-robot networks as highly reconfigurable systems for exploring non-equilibrium topological physics, while paving the way for topologically protected, robust collective behaviors in active matter.**

## INTRODUCTION

Topological phases of matter, celebrated for their ability to host robust boundary states immune to backscattering, were conceptualized within Hermitian frameworks that govern closed systems. [1-24]. While real-world systems are inherently open and dissipative, which necessitates non-Hermitian (NH) descriptions. In these NH descriptions, complex eigenvalue windings and exceptional point singularities give rise to unprecedented phenomena: parity-time symmetry transitions [25-27], dynamic encircling [28, 29], Weyl exceptional rings [30, 31], exceptional nexus [32], and NH topological modes [27, 33-38]. NH skin effects (the collapse of bulk states to the boundary) have been demonstrated in photonic [27, 39, 40], electrical [38], acoustic [33-35], mechanical [36, 37], and ultracold atomic [41, 42] systems. Recently, delocalized TZMs extended throughout bulks but not localized at boundaries have also been theoretically predicted [43, 44] and experimentally observed [36]. Those breakthroughs on NH topologies have spurred technological innovations, spanning single-mode lasing [45], light steering [40], to enhanced sensing [46, 47].

Despite these advances, a fundamental constraint remains: the implementation of NH topology has been largely confined to wave and matter systems [26, 27, 33-41, 48-53]. Although active metamaterials can digitally synthesize non-reciprocal feedback[36, 53, 54], their underlying couplings remain restricted to passive elastic beams. This stands in stark contrast to biological swarms (e.g., avian flocks or insect swarms) whose collective motions (e.g., velocity alignment, directional synchronization) emerge from local behavioral responses and real-time state exchanges [55]. Although recent robotic experiments have demonstrated non-reciprocal phase transitions by imitating such biological swarms [56], it remains an open question whether NH topologies can be emerged in collective intelligence systems unconstrained by spatial dimensionality.

Here, we address this challenge by experimentally demonstrating NH topological states within multi-robot networks. By digitally programming non-reciprocal behavioral rules and enabling real-time state exchange, we observe both localized and delocalized TZMs, alongside NH skin effects, across synthetic lattices spanning one to three dimensions. Unlike conventional wave-based systems tethered to holistic, global wavefield coherence, the observed non-Hermitian topological states are an emergent property of localized interaction dynamics within decentralized robotic networks. This architecture bypasses the requirement for centralized coordination, establishing a synthetic counterpart to natural self-organizing systems. Thus, this decentralized mechanism, previously unattainable in traditional physical platforms, unlocks a highly reconfigurable and scalable framework for exploring high-dimensional NH physics.

## RESULTS

### 1D NH multi-robot network

As shown in Fig. 1a, the robot contains three core components: an actuation motor, a sensor-control module, and a communication-computation module. Each robot continuously monitors its rotational motion state through an integrated position sensor and exchanges angular displacement information with its nearest neighbors via a communication unit. The multi-robot network follows a Su-Schrieffer-Heeger (SSH) model (Fig. 1b), where each unit (indexed by $a$, $a$=1, 2, …) contains two robots: the odd-indexed one #($2a-1$) and the even-indexed one #($2a$). The torques of robots follow two relationships. (1) Non-reciprocal hopping. Torques $M_{2a-1}^{1}$ and $M_{2a}^{1}$ of robots #($2a-1$) and #($2a$) are generated through the linear angular dependence on $\theta_{2a}$ and $\theta_{2a-1}$: $M_{2a-1}^{1} = -\delta\theta_{2a}$ and $M_{2a}^{1} = \delta\theta_{2a-1}$. $\delta$ denotes the non-reciprocal hopping constant. $\theta_{2a-1}$ and $\theta_{2a}$ are angular displacement of robots #($2a-1$) and #($2a$) in the unit cell ($a$). (2) Reciprocal interactions. Spring-equivalent stiffness coefficients ($k_1$, $k_2$) produce reciprocal torques: $M_{2a-1}^{2} = k_1(\theta_{2a} - \theta_{2a-1}) + k_2(\theta_{2a-2} - \theta_{2a-1})$ and $M_{2a}^{2} = k_1(\theta_{2a-1} - \theta_{2a}) + k_2(\theta_{2a+1} - \theta_{2a})$. $\theta_{2a-2}$ and $\theta_{2a+1}$ are angular displacements of robots #($2a-2$) and #($2a+1$) in adjacent unit cells ($a\mp1$). Total torques of robots #($2a-1$) and #($2a$) are $M_{2a-1}^{total} = M_{2a-1}^{1} + M_{2a-1}^{2}$ and $M_{2a}^{total} = M_{2a}^{1} + M_{2a}^{2}$ (Supplementary Information S1). This multi-robot network yields two critical features. (1) Equivalent on-site potential $-(k_1+k_2)$ of self-energy terms: $M'_{2a-1} = -(k_1 + k_2)\theta_{2a-1}$ and $M'_{2a} = -(k_1 + k_2)\theta_{2a}$. (2) Asymmetric interaction of distinct cross-terms: $M''_{2a-1} = (k_1 - \delta)\theta_{2a}$ and $M''_{2a} = (k_1 + \delta)\theta_{2a-1}$. In contrast to conventional non-reciprocal systems with physical linkages, our approach eliminates mechanical coupling components (such as springs or interconnects) through preprogrammed coupling protocols embedded in control chips. Then, exchanging angular displacement information between adjacent robots in real-time, we effectively construct equivalent dynamical relationships. Each robotic agent autonomously regulates its local dynamics governed by the programmed coupling rules, its own instantaneous kinematic state, and the real-time state feedback of adjacent units. Rather than relying on a centralized, global coordinator, this decentralized mechanism elegantly recapitulates the emergence of complex, robust group-level behaviors from simple, localized interactions.

Mechanical damping of the motor induces energy dissipation, which may degrade rotational motion precision. To mitigate damping effects, we reconfigure the motion control architecture by replacing conventional torque-based actuation with a motor angle control strategy. The angular displacement of the robot during the $j$-th time step follows $\Delta\theta_j = \left(\frac{M^{total}}{I}T + 2\omega_{j-1}\right)T/2$, where $\omega_{j-1}$ represents the angular velocity at the ($j-1$)-th step, $T$ is the time-step duration, and $I$ is the moment of inertia. We then achieve a multi-robot network exhibiting NH topological dynamics. The system features an on-site potential term given by $k_{1D} = k_1 + k_2$, and the TZM eigenfrequency is given by $f_1 = \frac{1}{2\pi}\sqrt{k_{1D}/I}$, where the moment of inertia $I$=1 kg·m$^2$. The square of TZM eigenfrequency exhibits a linear dependence on parameters $k_1$ and $k_2$, and is independent of the non-reciprocal transition constant $\delta$. The total number of robots is odd, and this framework enables programmable SSH-type collective dynamics, manifesting both delocalized and localized TZMs (Supplementary Information S1).

The digital encoding of coupling can naturally extend to full robotic digitization. Digital counterparts of robots autonomously capture rotational states, enable distributed angle information exchange, and compute subsequent rotation motion through the predefined coupling mechanism. As demarcated in Fig. 1c, physical robots (solid boxes) interface with digital ones (dashed boxes) through a bidirectional communication protocol. The latest physical robot $\#(2l-1)$ and the first digital robot $\#(2l)$ comprise the $l$-th unit cell. The angular displacement $\theta_{2l-1}$ of the physical robot $\#(2l-1)$ modulates the rotation of the digital robot $\#(2l)$ through the coupling kernel $M''_{2l} = (k_1 + \delta)\theta_{2l-1}$, and gradually determines collective motion patterns of digital robots. Concurrently, the angular information $\theta_{2l}$ of the digital robot $\#(2l)$ feeds back to the physical robot $\#(2l-1)$ via the coupling kernel $M''_{2l-1} = (k_1 - \delta)\theta_{2l}$. This reveals a physical-digital duality, the fusion phenomenon where both physical and digital robots exhibit isomorphic dynamics.

We first construct a hybrid multi-robot network (HMRN) consisting of 11 physical robots (indexed 1-11, left-right spatial encoding) and 90 digital counterparts to study the NH topological dynamics of non-reciprocal collective intelligence architectures. The intercellular (and extracellular) coupling stiffness is $k_1$=0.8 N·m/rad (and $k_2$=1 N·m/rad). The diagonal elements of the dynamic matrix formulation are $k_1$+$k_2$. Robot #1 is driven at the TZM eigenfrequency $f_1 \approx 0.21$ Hz. The dynamics of multi-robot network are systematically investigated under two distinct non-reciprocal hopping regimes. For the non-reciprocal hopping $\delta$=0.2 N·m/rad, the theoretical proportion coefficient between adjacent odd-indexed robots is $\alpha = -1$ (Supplementary Information S1). This configuration predicts a delocalized TZM propagation, experimentally validated by the motion state of odd-indexed robots from #1 to #11. Time-resolved angular amplitudes reveal two distinct dynamical phenomena (Fig. 2a, c and Supplementary Video S1). (1) Odd-indexed robots exhibit active propagations characterized by uniform angular displacements ($|\alpha| =$1.005±0.046), $\pi$-phase alternations between adjacent ones, and linear amplitude growths $|\theta(t)| \propto t$ ($|r| = 0.9996$, p<0.001) consistent with undamped forced vibration theory (Supplementary Information S4). (2) Even-indexed robots exhibit quasi-stationary states. When $\delta = -0.2$ N·m/rad, the amplitude ratio is $\alpha = -0.6$ (Supplementary Information S1), and the NH topologies of the non-reciprocal system transition to the localized TZM. Odd-indexed robotic measurements demonstrate exponential decay ($|\alpha| =$ 0.600±0.035), while even-indexed units also maintain quiescence (Fig. 2b, d and Supplementary Video S1).

Removing 90 digital robots, HMRN is degenerated to a physical multi-robot network (PMRN) consisting of only 11 physical robots (Supplementary Information S5). As shown in Fig. 2e and f, for both delocalized and localized TZMs, mean values of angular displacement ratios of PMRN and HMRN are close to theoretical values. We find that PMRN exhibit progressive error accumulation with positional displacement, where standard deviations of $|\alpha|$ escalate from 0.078 (or 0.038) of robot #3 to 0.269 (or 0.132) of robot #11 for delocalized (or localized) TZMs, as shown in Fig. 2e and f. This error phenomenon is effectively mitigated in HMRN through digital robot supplementation (Fig. 2e, f, and Supplementary Table S3). In addition, we plot violin statistics for quantifying distribution characteristics of $|\alpha|$, as shown in Fig. 2g and h. For delocalized (or left-localized) TZMs, HMRN increases the probability density of $|\alpha|$ in the interval [0.95, 1.04] (or [0.55, 0.64]) from 35.85% (57.58%) to 76.33% (85.39%), compared to PMRN. Therefore, digital robots, by increasing the length of the robot chain, make dynamic behaviors of physical robots convergent toward theoretical models, and finally enhance their motion robustness.

Notably, the periodic excitation at the frequency 0.53 Hz, offsetting from the TZM eigenfrequency (0.21 Hz), eliminates TZM responses, triggering NH skin effects (NHSEs) characterized by $\delta$-dependent spatial asymmetry (Supplementary Video S2): right-localized NHSEs ($\delta$ =0.2 N·m/rad, energy density is localized in the rightmost boundary, Fig. 2i) and left-localized NHSEs ($\delta = -0.2$ N·m/rad, energy density is localized in the leftmost boundary, Fig. 2j), with topological phase transitions at $\delta$=0.

**2D NH multi-robot network**

We implement a 2D NH multi-robot network on a square lattice with a flux quantization ($\pi$). It comprises 51×51 robots, including 5×5 physical robots and 2576 digital counterparts (Fig. 3a, b). They are governed by embedded coupling protocols. The defined couplings between robots are shown in Fig. 3a and Supplementary Information S2. Spring-equivalent stiffness coefficients are $k_1^x = k_1^y = 0.6$ N·m/rad and $k_2^x = k_2^y = 0.8$ N·m/rad. The diagonal elements of dynamic matrix are on-site potential terms, $k_{2D} = k_1^x + k_1^y + k_2^x + k_2^y$ . This causes a shift of the eigenvalue, when compared with the tight-binding model. The TZM eigenfrequency is $f_2 = \frac{1}{2\pi}\sqrt{k_{2D}/I}$ (Supplementary Information S2). The moment of inertia is $I$=1 kg·m$^2$. The TZM eigenfrequency is independent of non-reciprocal coupling constants $\delta_x$ and $\delta_y$.

We systematically investigate TZM responses in this 2D multi-robot network under a periodic excitation 0.27 Hz (the TZM eigenfrequency). Apply this stimulus to robot #(1, 1). When $\delta_x$=0.2 N·m/rad and $\delta_y$=0.2 N·m/rad, the amplitude ratios along the $x$ and $y$ directions are $\alpha_x = \alpha_y = -1$, theoretically predicting delocalized TZMs (Supplementary Information S2). As shown in Fig. 3c, experimental measurements demonstrate uniform angular amplitudes ($|\alpha_x| = 0.955 \pm 0.070$ and $|\alpha_y| = 1.016 \pm 0.148$) in dual-odd-indexed robots (method). When $\delta_x = -0.2$ N·m/rad and $\delta_y = 0.2$ N·m/rad (Fig. 3d), anisotropic amplitude ratios [$\alpha_x = -0.5$ (exponential decay) and $\alpha_y = -1$ (dislocalized propagation)] induce edge-localized TZMs (Supplementary Information S2). Measured rotations are mainly localized to robots #(1, 1), # (1, 3), and # (1, 5) along the $x$=1 edge, with uniform amplitudes ($|\alpha_y| = 0.984 \pm 0.038$). Away from the $x$=1 edge, the dual-odd-indexed robots (such as robots #(1, 1), # (3, 1), and # (5, 1)) experience exponential decay ($|\alpha_x|$ =0.521±0.047). When $\delta_x = -0.2$ N·m/rad and $\delta_y = -0.2$ N·m/rad(Fig. 3e), the bidirectional exponential decay ($\alpha_x = \alpha_y = -0.5$) confines the rotational energy to robot #(1, 1) (Supplementary Information S2). Experimental results show amplitude attenuation in both $x$ and $y$ direction ($|\alpha_x| = 0.475 \pm 0.075$ and $|\alpha_y| = 0.449 \pm 0.048$). For the above three cases, robots with non-dual-odd indexes exhibit amplitude suppression (Supplementary Video S3).

We then construct a 2D NH multi-robot network consisting of 5×5 physical robots by removing digital counterparts. Delocalized, edge-localized, and corner-localized TZMs are experimentally observed under the TZM eigenfrequency 0.27 Hz. Comparing with the larger-scale system with digital robots, this physical multi-robot network exhibits a compromised control precision and a reduced stabilization capacity (Supplementary Information S6 and Supplementary Table S4). When the driving frequency is 0.58 Hz (deviated from TZM eigenfrequency 0.27 Hz), TZMs undergo catastrophic collapse, activating $\delta$-dependent NHSEs with spatial energy localization at corners (Supplementary Video S4). Initially delocalized ($\delta_x$=0.2 N·m/rad and $\delta_y$=0.2 N·m/rad) and edge-localized ($\delta_x$=0.2 N·m/rad and $\delta_y = -0.2$ N·m/rad) TZMs under 0.58 Hz cascade into NH skin modes, respectively localized at corners (5, 5) and (1, 5), as shown in Fig. 3f-g. Corner-localized TZM at the robot # (1, 1) also transforms to the NH skin mode with a much larger attenuation (Fig. 3h).

### 3D NH multi-robot network

We demonstrate programmable TZMs in a 3D non-reciprocal multi-robot network. As shown in Fig. 4a, the cubic lattice with a uniform flux ($\pi$) on the square surface of the unit cell is adopted. The network comprises 15×15×15 robots, including 3×3×3 physical robots and 3348 digital ones. The multi-robot network is governed by non-reciprocal coupling relationship with equivalent stiffness parameters ( $k_1^x = k_1^y = k_1^z = 0.5$ N·m/rad and $k_2^x = k_2^y = k_2^z = 0.7$ N·m/rad). The diagonal elements of dynamic matrix are on-site potential terms $k_{3D} = k_1^x + k_1^y + k_1^z + k_2^x + k_2^y + k_2^z$. This causes a shift of the eigenvalue, when compared with the tight-binding model (Supplementary Information S3). The TZM eigenfrequency is $f_3 = \frac{1}{2\pi}\sqrt{k_{3D}/I}$. It is also independent of non-reciprocal coupling constants $\delta_x$, $\delta_y$, and $\delta_z$. Here, $I$=1 kg·m$^2$ and $f_3 \approx 0.3$ Hz. The non-reciprocal hopping parameters ($\delta_x$, $\delta_y$, and $\delta_z$) enable programmable configuration control of TZMs (Supplementary Information S3 and Supplementary Video S5). As shown in Fig. 4b, delocalized TZM ($\delta_x = \delta_y = \delta_z = 0.2$ N·m/rad) drives bulk periodic rotation patterns with uniform amplitude ($|\alpha_x|$=1.041±0.116, $|\alpha_y|$=1.026±0.111, and $|\alpha_z|$=1.040±0.086) (Method). Surface-localized TZM ($\delta_x = \delta_y = 0.2$ N·m/rad, and $\delta_z = -0.15$ N·m/rad) confines dynamics to triple-odd-indexed positions on the $z$=1 plane (Fig. 4c, $|\alpha_x|$=1.085±0.121 N·m/rad and $|\alpha_y|$=1.044±0.153 N·m/rad), while inducing axial exponential decay in the $z$ direction ($|\alpha_z|$=0.487±0.066). Edge-localized TZM (Fig. 4d, $\delta_x$=0.2 N·m/rad and $\delta_y$=$\delta_z$= −0.15 N·m/rad) restricts rotations along the $y$=$z$=1 edge ($|\alpha_x|$=0.989±0.094), decaying exponentially in both the $y$ and $z$ directions ($|\alpha_y|$=0.490±0.066 and $|\alpha_z|$=0.556±0.064). Corner-localized TZM ($\delta_x$=$\delta_y$=$\delta_z$= −0.15 N·m/rad) achieves a 3D confinement (Fig. 4e, $|\alpha_x|$=0.484±0.061, $|\alpha_y|$=0.522±0.057, and $|\alpha_z|$=0.521±0.054) at the grid point (1, 1, 1) (the lower-left corner). Notably, robots at non-triple-odd-indexed positions universally are rotation suppression.

We subsequently construct a fully physical 3×3×3 non-Hermitian robotic network by omitting the digital emulators. At a resonant eigenfrequency of 0.3 Hz, we experimentally resolve the emergence of delocalized, surface-localized, edge-localized, and corner-localized TZMs. Compared to the hybrid digital-physical architecture, this pure physical multi-robot implementation exhibits degraded control precision and limited stabilization performance (Supplementary Information S7 and Supplementary Table S5). When the driving frequency is 0.62 Hz deviated from the TZM eigenfrequency, the TZM modes vanish, accompanied by the emergence of $\delta$-dependent NHSE. Notably, delocalized, surface-localized, edge-localized, and corner-localized TZMs collapse into corner-localized NH skin modes, where rotation becomes three-dimensionally confined at corners (3,3,3), (3,3,1), (3,1,1), and (1,1,1) (Fig. 4f-i, and Supplementary Video S6).

## CONCLUSIONS

In summary, we have established programmable multi-robot networks as a highly versatile and physically unconstrained platform for non-Hermitian topological physics. Through tailoring non-reciprocal parameters, we realize both localized and delocalized TZMs across synthetic lattices spanning one to three dimensions. Crucially, the emergence of NH topological states is decoupled from traditional wave-based requirements for global field coherence. Instead, they arise fundamentally from the localized interactions and decentralized responses of individual agents. This architecture eschews the need for centralized coordination, instead harnessing emergent behaviors reminiscent of natural self-organizing systems. It will not only redefine the design space for non-Hermitian topology, accelerate the discovery of exotic, out-of-equilibrium topological phases, but also bridge quantum-like topological principles with macroscopic active systems, establishing a new paradigm for topologically protected, adaptive collective behaviors in dynamic and unpredictable environments.

**METHODS**

## The realization of multi-robot network

The driver of the robot is a brushless motor MS4005 with an integrated high-precision sensor-control module. The control and sensing precisions are 0.01°. We use three STM32F103C8T6 microcontroller units (MCUs). Among them, MCU 1 calculates the displacement of the robot through the defined coupling relationships and the torque-displacement formula. MCU 2 uses the UART communication protocol to transmit the rotation information measured in each cycle to adjacent robots. MCU 3 transmits the measured rotation information to the upper computer. The time step is 25 ms.

The digital robots are implemented in a dedicated microcontroller, an STM32F406RGT6 chip. It completes the angular displacement calculations of digital robots within a time step. It is also responsible for storing, exchanging, and computing the rotation information of digital robots. It should be noted that digital robots execute virtual movements instead of real rotations.

## TZMs and NHSEs

In the one-, two-, and three- dimensional multi-robot networks, the excitation torque is $M = M_0\sin(\omega t)$. We discretize the excitation torque $M = M_0\sin(\omega t)$. The time step $T$ is 25 ms. Then, the value of excitation torque at each moment is $M = M_0\sin(0.025\omega nT)$, where $n$=1, 2, .... The resultant torque of the excited robot at each cycle is the sum of the excitation torque $M$ and the additional torques determined by the excited robot and its neighboring ones (Supplementary Information S1, 2, 3, 4).

In the one-dimensional multi-robot network, TZMs are excited at the eigenfrequency $f = \frac{1}{2\pi}\sqrt{k_{1D}/I}$ (Supplementary Information S1). The excited robot is at the left point #1. In the two-dimensional multi-robot network, TZMs are excited at the eigenfrequency $f = \frac{1}{2\pi}\sqrt{k_{2D}/I}$ (Supplementary Information S2). The excited robot is at the bottom left point #(1, 1). In the three-dimensional multi-robot network, TZMs are excited at the eigenfrequency $f = \frac{1}{2\pi}\sqrt{k_{3D}/I}$ (Supplementary Information S3). The excited robot is at the point #(1, 1, 1). When the excited frequency deviates from the eigenfrequency, these TZMs transition to NH skin modes.

## Experiments of one-, two-, and three-dimensional multi-robot networks

In the one-dimensional multi-robot network described in the main text, the TZM responses from $t$ = 34.45s to 80s are recorded. $\alpha_a(t_j) = \theta_{2a-1}(t_j)/\theta_{2a-3}(t_j)$ represents the angular displacement ratios of the $a$-th unit cell, $a$=2, 3, 4, 5, 6. $\alpha = \sum_{a=2}^{6}\sum_{j=1}^{J}\alpha_a(t_j)/(5J)$ represents the experimental proportion coefficient of the one-dimensional multi-robot network. For robots situated at the boundaries of the multi-robot network, additional springs ($k_2$ on the left boundary and $k_1$ on the right boundary) are introduced to ensure the equality of diagonal elements of the dynamic matrix. Outer sides of additional springs are fixed. In Fig. 2i and j, $u$ represents the energy density of the robot. $u = \frac{1}{J}\sum_{j=1}^{J}\left(E_k(t_j) + E_p(t_j) + E_o(t_j)\right)$, where $E_k(t_j)$ represents the kinetic energy of the robot, $E_p(t_j)$ represents the spring potential energy between robots, $E_o(t_j)$ represents the potential energy possessed by the additional robot's equivalent spring, and $J$ represents the total number of time samples. $\Delta t_j = 0.05s$.

For the two-dimensional multi-robot network (comprising both physical and digital robots) described in the main text, the TZM responses from $t$ = 43.2s to 46.9s are recorded. As shown in Fig. 3c-e, responses of robots are normalized by the formula $\theta'_{x,y} = \theta_{x,y}/\theta_{1,1}$, $\theta_{x,y}$ represents angular displacement of the robot #$(x, y)$, and $\theta'_{x,y}$ represents the normalized value. $\alpha_{a,b}^{x}$ ($\alpha_{a,b}^{y}$) represents the angular displacement ratios of $a$-th ($b$-th) unit cell in the $x$ ($y$) direction. $\alpha_{a,b}^{x} = \theta_{2a-1,2b-1}/\theta_{2a-3,2b-1}$ ($a$=2-3 and $b$=1-3) and $\alpha_{a,b}^{y} = \theta_{2a-1,2b-1}/\theta_{2a-1,2b-3}$ ($a$=1-3 and $b$=2-3). $\theta_{2a-1,2b-1}$ is the angular displacement of dual-odd-indexed robot #(2$a$–1, 2$b$–1). $\alpha_x$ ($\alpha_y$) represents the experimental proportion coefficient of the multi-robot network in the $x$ ($y$) direction. $\alpha_x = \sum_{b=1}^{3}\sum_{a=2}^{3}\sum_{j=1}^{J}\alpha_{a,b}^{x}(t_j)/(6J)$ and $\alpha_y = \sum_{b=2}^{3}\sum_{a=1}^{3}\sum_{j=1}^{J}\alpha_{a,b}^{y}(t_j)/(6J)$. The energy density in Fig. 3f-h is expressed as $u = \frac{1}{J}\sum_{j=1}^{J}\left(E_k(t_j) + E_p(t_j) + E_o(t_j)\right)$. $\Delta t_j = 0.1$s. Similarly, the robots located at the boundary of the two-dimensional multi-robot network possess an additional potential energy term. The exact value of the additional potential energy term is determined in Part A of Supplementary Information S2.

In the three-dimensional multi-robot network, the TZM responses during $t$ = 55s to 65s are recorded. As shown in Fig. 4b-e, we normalize the responses of robots by $\theta'_{x,y,z} = \theta_{x,y,z}/\theta_{1,1,1}$. $\alpha_{a,b,c}^{x}$ ($\alpha_{a,b,c}^{y}$ and $\alpha_{a,b,c}^{z}$) represents the angular displacement ratio of the $a$-th ($b$-th and $c$-th) unit cell in the $x$ ($y$ and $z$) direction. $\alpha^{x}_{a,b,c}=\frac{\theta_{2a-1,2b-1,2c-1}}{\theta_{2a-3,2b-1,2c-1}}$ ($a$=2, $b$=1,2, and c=1,2), $\alpha^{y}_{a,b,c}=\frac{\theta_{2a-1,2b-1,2c-1}}{\theta_{2a-1,2b-3,2c-1}}$ ($a$=1,2, $b$=2, and c=1,2) , and $\alpha^{z}_{a,b,c}=\frac{\theta_{2a-1,2b-1,2c-1}}{\theta_{2a-1,2b-1,2c-3}}$ ($a$=1,2, $b$=1,2, and c=2). $\theta_{2a-1,2b-1,2c-1}$ is the angular displacement of treble-odd-indexed robot #(2$a$–1, 2$b$–1, 2c-1). $\alpha_x$ ($\alpha_y,\alpha_z$) represents the experimental proportion coefficient of the robots in the $x$ ($y$, $z$) direction. $\alpha_x=\sum_{c=1}^{2}\sum_{b=1}^{2}\sum_{j=1}^{J}\alpha^{x}_{2,b,c}(t_j)/(4J)$, $\alpha_y=\sum_{c=1}^{2}\sum_{a=1}^{2}\sum_{j=1}^{J}\alpha^{y}_{a,2,c}(t_j)/(4J)$, and $\alpha_z=\sum_{b=1}^{3}\sum_{a=1}^{3}\sum_{j=1}^{J}\alpha^{z}_{a,b,2}(t_j)/(4J)$. The energy density of NH skin effects in Fig. 4f-i is expressed as $u=\frac{1}{J}\sum_{j=1}^{J}\left(E_k(t_j)+E_p(t_j)+E_o(t_j)\right)$. $\Delta t_j=0.1$s. The sampling time for Fig. 4f-i is $t$=17.3-65 s.

**Acknowledgements:** This work was supported by the Natural Science Foundation of Hunan Province for Young Scientists Fund (Category A, including continuation funding projects) (Grant No. 2026JJ20001), the National Natural Science Foundation of China (Grant Nos. 52475254, 52505097, and 12072108), and the Project of State Key Laboratory of Advanced Design and Manufacturing for Vehicle Body.

**Author contributions:** J.Z. and B.X. designed the research. J.Z., G.D., and T.C. conducted theoretical analysis. J.Z., T.C., and B.X. conducted the measurements. J.Z., G.D., and T.C. analysed the data. J.Z., G.D., T.C., and B.X. wrote the paper. B.X. supervised the project.

**Competing interests:** The authors declare no competing interests.

**Data andmaterials availability:** All data needed to support the conclusions of this manuscript are included in the main text or Supplementary Materials. All the codes are available.

# Figures

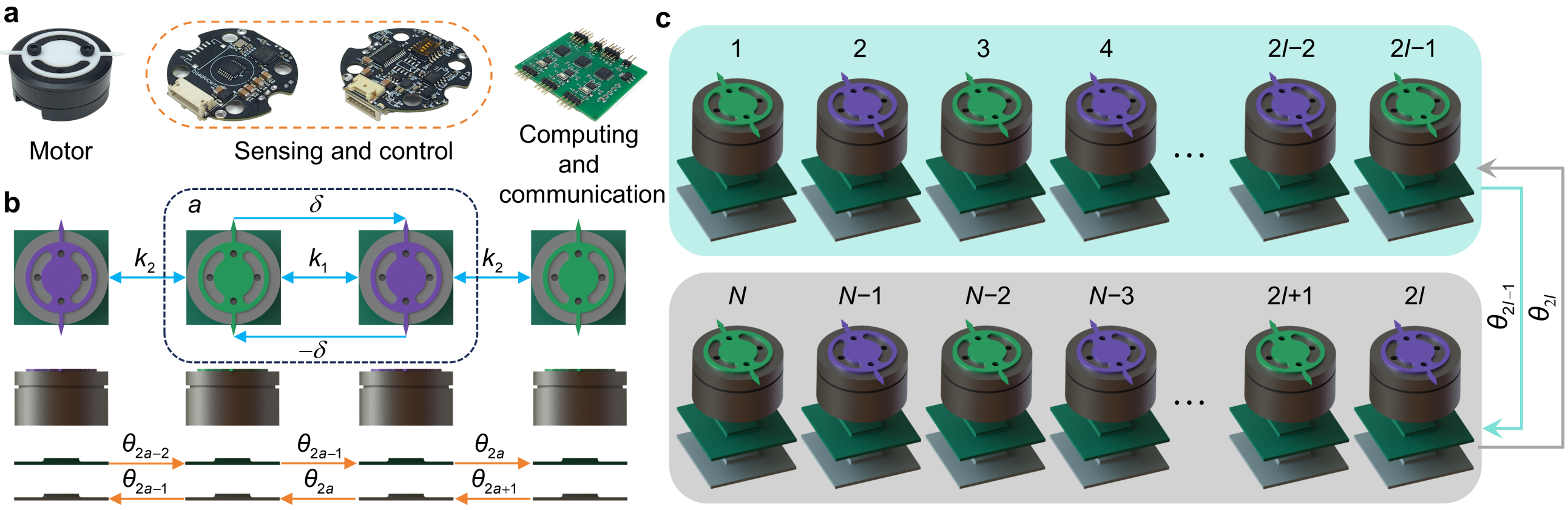

**Fig. 1. 1D NH multi-robot network. a**, The robot consists of the actuation motor, the sensor-control module, and the communication-computation module. **b**, Couplings between robots. The light blue lines with double arrows represent the reciprocal couplings ($k_1$ and $k_2$). The light blue line with an arrow represents the non-reciprocal hopping ($\delta$ or $-\delta$). The orange arrow represents the transmission of angular displacement information between robots. The two robots in the dashed box form a unit cell. **c**, The 1D NH multi-robot network with $N$ robots. Robots in the blue box represent physical ones, while the others in the gray box represent digital ones. Coupling and communicating protocols between digital robots are the same as those defined in (**b**). Coupling and communicating protocols between digital $\#(2l-1)$ and physical $\#(2l)$ robots are the same as those defined in (**b**).

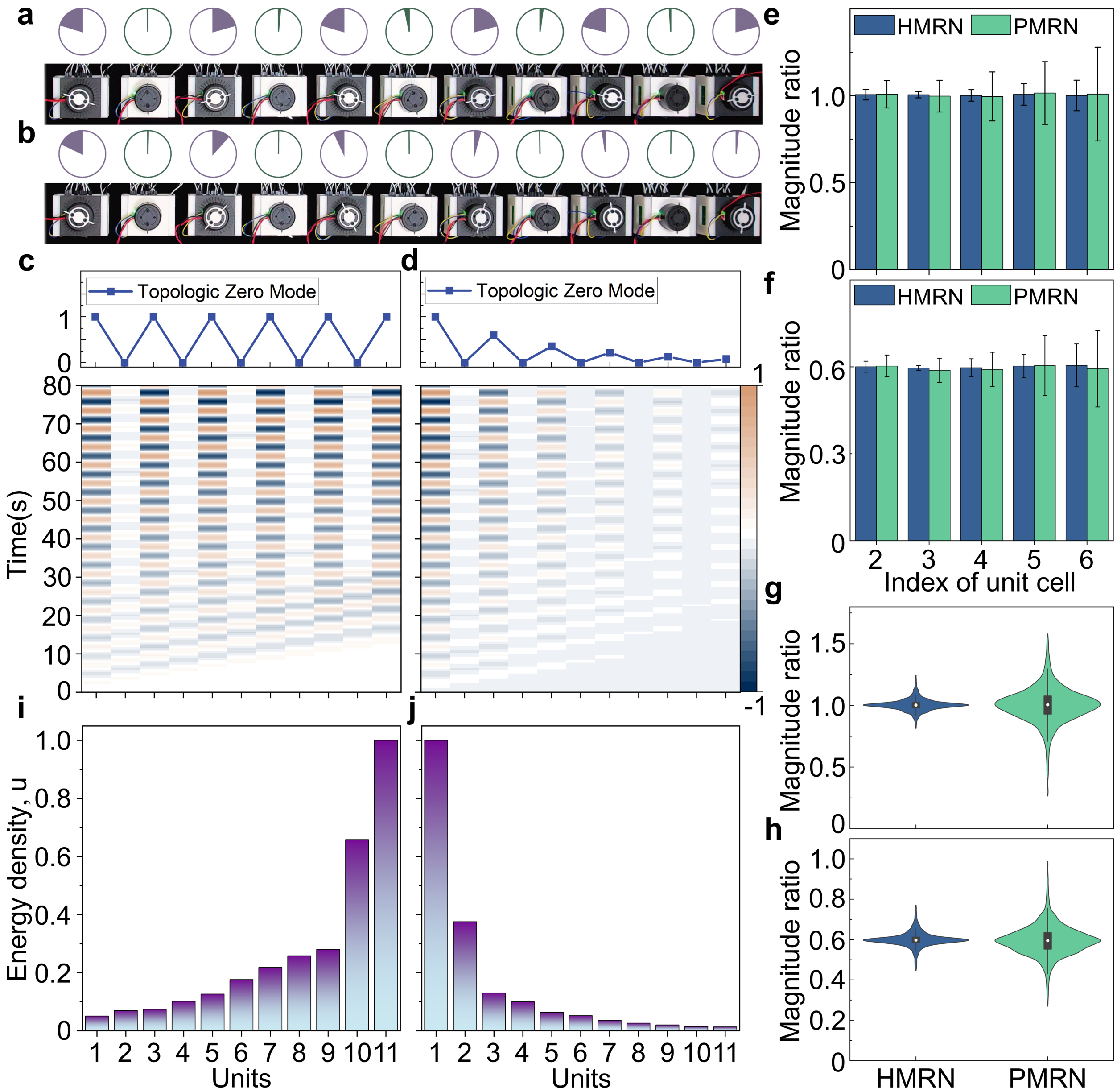

**Fig. 2. Topological zero modes and skin states of the 1D NH multi-robot network. a**, The delocalized TZM responses ($\delta = 0.2$ N·m/rad) of the 1D NH multi-robot network at a time $t$=49.85s. The rotation angles (rad) of robots #1 to #11 are 1.28, 0.02, –1.30, –0.09, 1.29, 0.20, –1.31, –0.14, 1.33, 0.06, –1.32. **b**, The left-localized TZM responses of the 1D NH multi-robot network at a time $t$=49.75s. The rotation angles (rad) of robots #1 to #11 are 1.21, –0.03, –0.73, –0.01, 0.44, 0.01, –0.26, 0.01, 0.16, –0.01, –0.09. **c** and **d**, Time-domain responses ($t$=0-80s) of the 1D NH multi-robot network with delocalized and left-localized TZMs. **a**, **b**, **c**, and **d**, The robot #1 is subjected to a periodic excitation of 0.21Hz. **e** and **f**, Means and standard variances of ratios $|\alpha|$ under delocalized (**e**) and left-localized (**f**) TZMs. **g** and **h**. Violin diagrams of set of ratios ($|\alpha|$) under delocalized and left-localized TZMs. Deep blue and green respectively represent hybrid multi-robot network (HMRN) and physical multi-robot network (PMRN). Digital robots significantly reduce standard variances and improve violins' concentrations. **i** and **j**, The energy densities of PMRN when subjected to a periodic excitation $f$=0.53 Hz, offsetting from the TZM eigenfrequency.

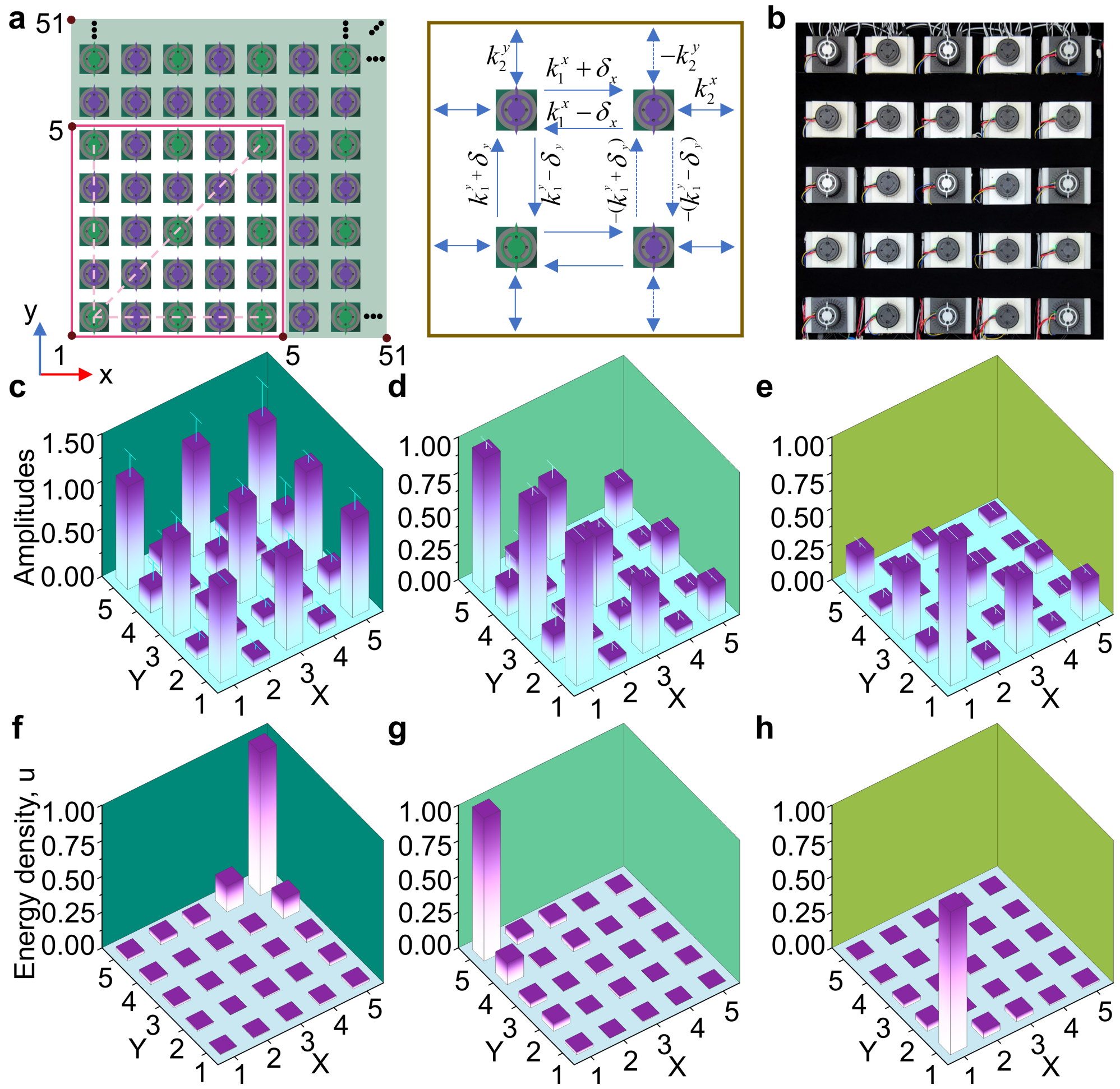

**Fig. 3. Topological zero modes and skin state of the 2D NH multi-robot network. a**, The 2D NH multi-robot network. Robots in the carmine solid line box are physical ones, while the others are digital ones. Four robots form a unit cell with a flux quantization ($\pi$) analogous to a 2D quadrupole model. The dashed line indicates the opposite coupling strength. **b**, Experimental diagram of a two-dimensional NH multi-robot network. There are 5×5 physical robots and 2576 digital robots. **c**, **d**, and **e**, The responses of 2D NH multi-robot network under the TZM eigenfrequency (0.27Hz). They exhibit delocalized, edge-localized, and corner-localized TZMs, respectively. **f**, **g**, and **h**, The energy densities of 2D NH multi-robot network under the frequency 0.58 Hz away from the TZM eigenfrequency. They exhibit corner-localized skin states whose locations are determined by $\delta_x$ and $\delta_y$.

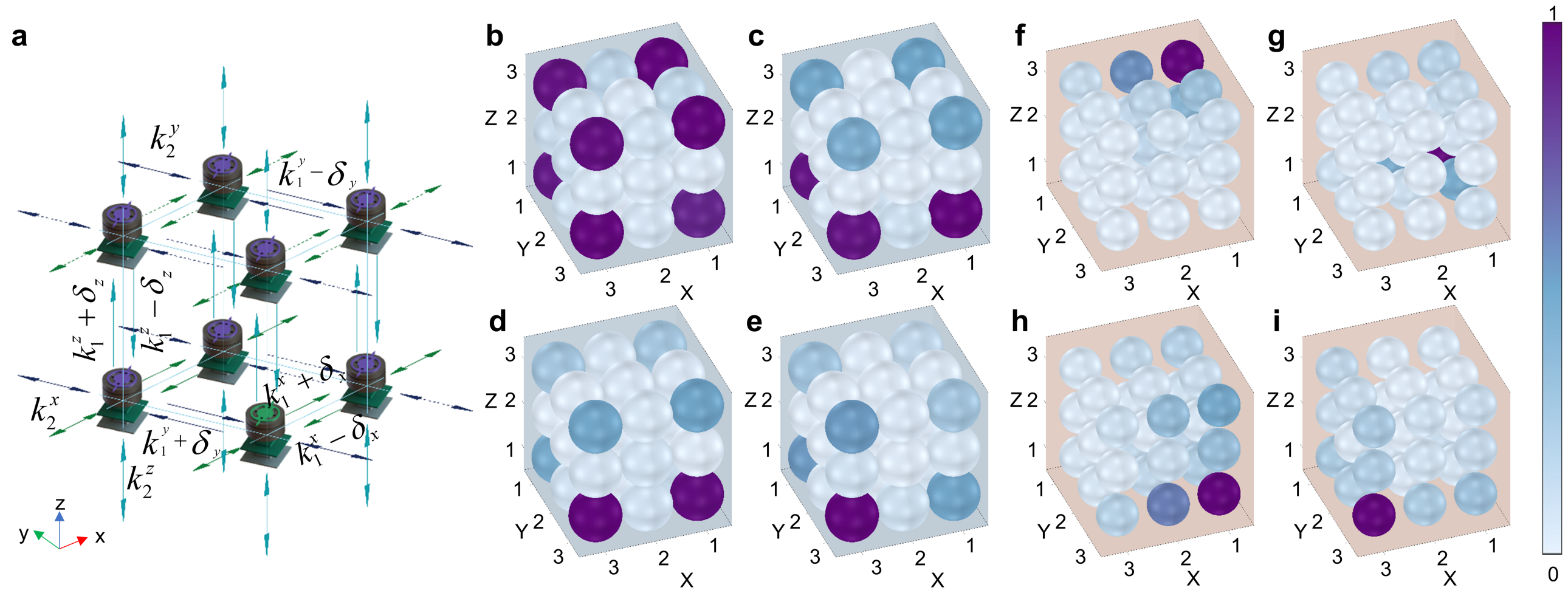

**Fig. 4. Topological zero modes and NH skin states of the 3D NH multi-robot network. a**, The 3D NH multi-robot network. Eight robots form a unit cell by analogous of a 3D octupole model. **b, c**, **d** and **e**, the responses of the 3D NH multi-robot network under the TZM eigenfrequency (0.3Hz). They exhibit delocalized, surface-localized, edge-localized, and corner-localized TZMs, respectively. **f**, **g**, **h** and **i**, The energy densities (Method) of the 3D NH multi-robot network under the frequency 0.62 Hz away from the TZM eigenfrequency. They exhibit corner-localized skin states whose locations are determined by $\delta_x$, $\delta_y$, and $\delta_z$.